\documentclass[prd,tightenlines,nofootinbib,prd]{revtex4}
\pdfoutput=1
\usepackage{amsmath,amscd,amssymb}
\usepackage{color,epsfig}
\usepackage{feynmf}
\usepackage{xfrac}
\usepackage{latexsym}
\usepackage{mathrsfs}
\usepackage{graphicx}
\usepackage{bbm}      % double-striked fonts for sets of integer numbers etc.

\newcommand{\imu}{{\rm i}}
\newcommand{\Vek}[1]{\mbox{\boldmath$#1$\unboldmath}}
\newcommand{\vek}[1]{\mathbf{#1}}

\begin{document}

\title{Vacuum polarization energy of a complex scalar field 
in a vortex background}

\author{N. Graham$^{a)}$, H. Weigel$^{b)}$}

\affiliation{
$^{a)}$Department of Physics, Middlebury College
Middlebury, VT 05753, USA\\
$^{b)}$Institute for Theoretical Physics, Physics Department, 
Stellenbosch University, Matieland 7602, South Africa}

\begin{abstract}
Scattering methods make it possible to compute the effects of
renormalized quantum fluctuations on classical field configurations.
As a classic example of a topologically nontrivial classical solution, the 
Abrikosov-Nielsen-Olesen vortex in U(1) Higgs-gauge theory provides an 
ideal case in which to apply these methods.  While physically measurable
gauge-invariant quantities are always well-behaved, the topological
properties of this solution give rise to singularities in gauge-variant
quantities used in the scattering problem.  In this paper we
show how modifications of the standard scattering approach are
necessary to maintain gauge invariance within a tractable calculation.
We apply this technique to the vortex energy calculation in a
simplified model, and show that to obtain accurate results requires an
unexpectedly extensive numerical calculation, beyond what has been
used in previous work.
\end{abstract}

\maketitle

\section{Introduction}

Topological vortex configurations appear in many field theory models
in the form of axially symmetric $U(1)$ vector potentials pointing in
azimuthal direction.  Although physically measurable quantities remain
well-defined, in these models the field profiles, which depend on the
gauge choice, necessarily have a singular  structure. This structure
hampers many computations that go beyond the (static) mean-field approach,
particularly the quantum correction to the classical static energy density  
per unit length. At one loop order this quantity is given by the vacuum 
polarization energy (VPE), the renormalized shift in the sum of the zero 
point energies of the quantum fluctuations. Calculations of the VPE often use
auxiliary quantities like Feynman diagrams and/or expansion schemes for 
scattering data \cite{Graham:2009zz}. These quantities are  not necessarily 
gauge invariant and thus the singular structure matters.

The classic example of a topological string solution is the Abrikosov-Nielsen-Olesen (ANO)
vortex \cite{Abrikosov:1956sx,Nielsen:1973cs}. In this paper we use a simplified model that 
contains all the relevant subtleties and obstacles of the full Nielsen-Olesen 
model, but is computationally simpler. As in the full model, we consider a 
complex scalar field (representing a Higgs boson in particle physics or a
Cooper pair condensate in condensed matter physics) coupled to
the U(1) gauge field of electromagnetism.  However, we consider only
fluctuations of the scalar field; adding the gauge field fluctuations
requires a more elaborate calculation, but will follow the same
formalism \cite{Graham:2019fzo}.

The quantum fluctuations propagate in a combined potential generated by both
the gauge and scalar field backgrounds. Typically the latter provides the 
dominant contribution, but the correct identification and renormalization of 
the ultra-violet divergences requires a detailed analysis of the interaction 
with the vortex background.

To our knowledge the only full calculation of the renormalized quantum
correction in the full Nielsen-Olesen model, as is made possible by the 
approach we will describe, is that of Refs.~\cite{Baacke:2008zx,Baacke:2008sq}.
Our approach will differ in several interconnected ways:  We carry out the 
calculation directly for the total energy, rather than as an integral over 
the energy density; we use analytic continuation
to carry out the integral over the wave number $k$ on the imaginary axis; and, 
most importantly, we express the calculation in a way that avoids 
superficial divergences originating from the singular gauge field profile. 
This process eliminates the need for the arbitrary additional interaction 
vertex that was added (and, correspondingly, subtracted back out) in 
Refs.~\cite{Baacke:2008zx,Baacke:2008sq}.  We will see that formidable
numerical challenges remain even after these simplifications, which 
suggests that the previous results are numerically
unreliable within a realistic computational budget.  Our analytic
continuation approach also sheds light on a puzzle first exhibited in
Ref.~\cite{Pasipoularides:2000gg}, which showed that calculations 
of quantum effects in a vortex background appear artificially
convergent when computed on the real $k$ axis; by carrying out the
computation in the complex $k$ plane from the outset, we will see how
the appropriate divergences emerge from the analytic structure of the
scattering data associated with the quantum fluctuations.

The complications that our method is capable of addressing can be
particularly acute in heat kernel methods, which are typically combined with
$\zeta$-function regularization, that have also been applied to vortex
systems~\cite{AlonsoIzquierdo:2004ru,Guilarte:2009hh,Alonso-Izquierdo:2016bqf}.
Not only do these methods involve complicated recursion relations for
the heat kernel coefficients that need to be truncated, but
the $\zeta$-function regularization also makes the implementation of
standard, perturbative renormalization conditions difficult up to
the point that gauge variant counterterms are
required~\cite{AlonsoIzquierdo:2004ru}. For certain geometries,
however, these methods have analytic
solutions~\cite{Ferreiros:2014mca}.

This paper is structured in eight short Sections. Following this introduction,
we introduce the model in Section II and collect the relevant ingredients 
to compute the VPE using spectral methods for configurations that are 
translationally invariant in (at least) one spatial variable in Section 
III. In Section IV we apply this method to regular configurations 
that couple to a (complex) scalar quantum field. In Section V we show how the 
ultra-violet divergences of Feynman diagrams relate to the asymptotic behavior 
of the scattering data that enter the VPE calculation. This analysis ensures 
the correct identification of the ultra-violet divergences on the scattering 
side of the calculation. We also discuss obstacles to computing the Feynman 
diagrams when the background configuration is singular. Section VI gives the 
main analytic calculation: We show how to modify the standard approach for
singular configurations for which the ingredients of the standard approach 
are not well defined.  In Section VII we perform numerical experiments for 
this modification and demonstrate that the numerical simulation requires a 
sophisticated computation to match the asymptotic behavior of the scattering 
data as given by the analysis of the Feynman diagrams. Finally, in Section VIII 
we summarize our findings and provide an outlook on how they affect the VPE 
calculation in the full model that also includes gauge field loops. 

\section{The model}

In rescaled variables, the Lagrangian for this model reads
\begin{equation}
\mathcal{L}=\left|\left(\partial_\mu-\imu A_\mu\right)\phi\right|^2
-\frac{m^2}{4}\left(|\phi|^2-1\right)^2\,.
\label{eq:lag}\end{equation}
The background vortex is translationally invariant along an axis that
we choose to be the $z$-axis. In terms of polar coordinates
$\rho$ and $\varphi$ in the  perpendicular plane, the vortex is
characterized by two radial functions:
\begin{equation}
\phi_0=f_H(\rho){\rm e}^{\imu \varphi}
\qquad {\rm and}\qquad \Vek{A}=-\hat{\Vek{\varphi}}\frac{f_G(\rho)}{\rho}\,.
\label{eq:vortex}\end{equation}
The ANO string \cite{Abrikosov:1956sx,Nielsen:1973cs,deVega:1976xbp} 
emerges as the stationary  solution when $\mathcal{L}$ is supplemented
by the gauge kinetic term $-(1/4)(\partial_\mu A_\nu-\partial_\nu
A_\mu)^2$. The boundary conditions  for the classical scalar field in
the vortex configuration are always given by
$\lim_{\rho\to0}f_H(\rho)=0$ and  $\lim_{\rho\to\infty}f_H(\rho)=1$. A
non-trivial topology requires
$\lim_{\rho\to0}f_G(\rho)-\lim_{\rho\to\infty}f_G(\rho)\ne0$, 
but the specific values depend on the choice of gauge. Calculating the
VPE will require the Fourier transforms of the classical vortex field
configurations.  In particular for the vector potential we have
\begin{equation}
\widetilde{\Vek{A}}(\Vek{p})=\int d^2x\, {\rm e}^{\imu\vek{p}\cdot\vek{x}}
\Vek{A}(\Vek{x})=2\pi\imu \hat{\Vek{\varphi}}_p
\int_0^\infty d\rho\, f_G(\rho)J_1(\rho p)\,,
\label{eq:FT1}\end{equation}
where $\hat{\Vek{\varphi}}_p$ is the azimuthal unit vector in momentum
space and $J_1(z)$ is a Bessel function. The integral on the right
hand side is well-defined only when
$\lim_{\rho\to\infty}f_G(\rho)=0$. Hence computing the VPE requires us
to choose the so-called {\it singular gauge} with
$\lim_{\rho\to\infty}f_G(\rho)=0$, meaning that
$\lim_{\rho\to0}f_G(\rho)\ne0$ for a configuration with nontrivial
topology. In Section \ref{sec:FT} we will actually see that even with
this particular condition on the gauge profile, special care is needed when
using Eq.~(\ref{eq:FT1}) to compute Feynman diagrams.

We introduce small amplitude fluctuations about the classical scalar field via
$\phi=\phi_0+\eta$ and derive their field equations in the harmonic
approximation,
\begin{equation}
\left[\partial_t^2-\nabla^2
+\frac{2\imu}{\rho^2}f_G(\rho)\partial_\varphi
+\frac{1}{\rho^2}f^2_G(\rho)-\frac{3m^2}{2}f_H^2(\rho)+\frac{m^2}{2}
\right]\eta=0\,.
\label{eq:flct1}\end{equation}
The partial wave-decomposition 
$\eta={\rm e}^{-\imu \omega t}\sum_\ell \eta_\ell(\rho)
{\rm e}^{\imu\ell\varphi}$ yields
\begin{equation}
\left[-k^2-\frac{1}{\rho}\partial_\rho \rho\partial_\rho
+\frac{\ell^2}{\rho^2}-\frac{2\ell}{\rho^2}f_G(\rho)
+\frac{1}{\rho^2}f^2_G(\rho)+V_H(\rho)\right]\eta_\ell(\rho)=0,
\label{eq:flct2}\end{equation}
with the dispersion relation $\omega^2=k^2+m^2$ and the abbreviation
$V_H(\rho)=\frac{3m^2}{2}(f_H^2(\rho)-1)$ for the scalar potential. We
see that the singular gauge boundary conditions described above turn 
Eq.~(\ref{eq:flct2}) into a free radial differential equation at spatial 
infinity. As a result, we can see explicitly that the 
Green's function constructed from $\eta_\ell$ will asymptotically approach 
its non-interacting counterpart so that the  vacuum energy density associated 
with the fluctuations vanishes at spatial infinity. This is another reason 
for choosing the singular gauge.

\section{Scattering data and vacuum polarization energy}

Eq.~(\ref{eq:flct2}) defines a potential scattering problem for which
we can obtain phase shifts and bound state energies in each partial
wave. Spectral methods use these scattering data to compute the VPE
\cite{Graham:2009zz}. In this approach, we first write the vacuum
expectation value of the energy density  operator in the presence of
the background fields in terms of the Green's function, which we can
then write as a sum over the full set of solutions to Eq.~(\ref{eq:flct2}).  
After subtracting the analogous contribution without the background and 
integrating over space, we obtain an expression for the total energy in 
terms of an integral over wave number $k$ and a sum over partial waves 
$\ell$. In this expression, the essential element, which encodes the 
dynamical information, is  the momentum derivative of the logarithm of the 
Jost function. For real momenta, this result can be expressed in terms of 
the phase of the Jost function, {\it i.e.} the phase shift \cite{Graham:2002xq}. 
For numerical purposes, this momentum derivative is typically treated via
integrating by parts. Our calculation will take advantage of the freedom to 
add an arbitrary constant to this logarithm without changing the VPE 
in order to maintain its analytic properties. In particular, we will show 
that by introducing a uniquely determined constant, which does not affect 
the calculation on the real $k$ axis, we are able to calculate the integral 
by continuing to imaginary wave number $t=\imu k$. There are then two 
ingredients to the integral. One arises from the  discontinuity of 
${\rm ln}\left(m^2-t^2\right)$ and the other from the poles of the
logarithmic derivative. These poles emerge at the bound state wave numbers, 
thereby canceling the bound state contribution to the VPE \cite{Bordag:1996fv}. 
Furthermore, on the imaginary axis functions that oscillate for real 
momenta \cite{Pasipoularides:2000gg} are replaced by their exponentially 
damped counterparts, making the calculation numerically more efficient.  
Finally, as we will see below, it is on the imaginary axis that we 
are able to recognize the full divergence structure of the theory.

The interface formalism \cite{Graham:2001dy} provides an
extension of the scattering approach to compute the VPE in cases
where the background potential is translationally invariant in
one or more spatial coordinates. To implement this formalism 
in our Higgs-gauge vortex model, we first define
\begin{equation}
\left[\nu(t)\right]_n=\lim_{L\to\infty} \sum_{\ell=-L}^{L}
\left[\nu_\ell(t)\right]_n\,,
\label{eq:jost1}\end{equation}
where $\nu_\ell(t)$ is the logarithm of the Jost function associated 
with orbital angular momentum $\ell$. The subscript $n$ denotes a suitable 
subtraction, typically in form of the Born approximation, such
that integral in 
\begin{equation}
E_{\rm VPE}=\frac{1}{2\pi}\int_m^\infty tdt\, \left[\nu(t)\right]_n
+E^{(n)}_{\rm FD}+E_{\rm CT}
\label{eq:vpe1}\end{equation}
is finite. The lower limit of the integral is the mass of the quantum
fluctuations, which for the current conventions is the scalar mass $m$, 
and the integral runs along the branch cut discontinuity mentioned above.
The subtractions in Eqs.~(\ref{eq:jost1}) and~(\ref{eq:vpe1}) are added 
back as (dimensionally) regularized Feynman diagrams $E^{(n)}_{\rm FD}$;
when combined with the counterterm  contributions ($E_{\rm CT}$), the
regulator may be removed in a renormalizable theory.  In the current
study we are primarily interested in the structure of
$\left[\nu(t)\right]_n$, in particular with regard to a singular
vortex background. Therefore we leave the details of 
$E^{(n)}_{\rm FD}$ and $E_{\rm CT}$ with on-shell renormalization
conditions~\cite{Irges:2017ztc} to a separate publication~\cite{Graham:2019fzo}, 
which computes the VPE in the complete Nielsen-Olesen model.

To numerically compute $\nu_\ell(t)$, we first write
$\eta_\ell(\rho)=K_{|\ell|}(t\rho)\overline{\eta}_\ell(\rho)$, where
$K_\ell$ is a modified Bessel function. This produces the differential
equation
\begin{equation}
\frac{1}{\rho}\partial_\rho \rho\partial_\rho\overline{\eta}_\ell=
2tZ_\ell(t\rho)\partial_\rho \overline{\eta}_\ell
+\frac{1}{\rho^2}\left[f_G^2-2\ell f_G\right]\overline{\eta}_\ell
+V_H\overline{\eta}_\ell
\qquad {\rm with}\qquad
Z_\ell(z)=\frac{K_{|\ell|+1}(z)}{K_{|\ell|}(z)}-\frac{|\ell|}{z}\,.
\label{eq:jost2}\end{equation}
We then integrate this differential equation with the boundary conditions
$\lim_{\rho\to\infty}\overline{\eta}_\ell(\rho)=1$ and 
$\lim_{\rho\to\infty}\frac{d}{d\rho}\overline{\eta}_\ell(\rho)=0$, and extract 
$\nu_\ell(t)=\lim_{\rho\to0}\ln\left(\overline{\eta}_\ell (\rho)\right)$.
The  Born series, which is the standard tool to remove the divergences in 
Eqs.~(\ref{eq:jost1}) and~(\ref{eq:vpe1}) \cite{Graham:2009zz}, expands the
solutions of Eq.~(\ref{eq:jost2}) in powers of the strength of the background 
potential $V_B=\frac{1}{\rho^2}\left[f_G^2-2\ell f_G\right]+V_H$. 
This expansion produces a series that approximates the scattering data 
well at large momenta. Hence subtracting the leading terms of this series from 
$\nu(t)$ renders the integral in Eq.~(\ref{eq:vpe1}) finite.

In contrast to the phase shift, which is the imaginary part of
$\nu_\ell$ for real momenta, the Jost function on the imaginary
momentum axis is not a gauge invariant quantity. Hence we can not rely
on gauge invariance when investigating the divergence structure
associated with $\nu_\ell$.

\section{Standard treatment for regular vector potentials}
\label{sec:standard}
\begin{fmffile}{quad}

Before addressing singular configurations, we review the
standard treatment for a regular gauge profile that vanishes at the
origin and spatial infinity, in order to properly identify the quantum
field theory divergences in Eq.~(\ref{eq:vpe1}).  Because
there are cancellations of divergences between various contributions
at different orders of $V_B$, it is appropriate to consider insertions of
$f_G^2/\rho^2$, $-2\ell f_G/\rho^2$ and $V_H$ separately. The first two of 
these terms originate from the $A_\mu A^\mu$ and $A_\mu\partial^\mu$
interactions, respectively,\footnote{Note that we work in a gauge where the 
vortex profile obeys $\partial_\mu A^\mu=0$.} in Eq.~(\ref{eq:lag}) and 
correspond to vertices with one or two external photon lines, while the 
third corresponds to vertices with one or two external scalar lines. 
The full VPE can then be written as the sum of one-loop Feynman diagrams 
with (the Fourier transforms of) the profiles emerging as external lines, 
combined with the corresponding contributions of renormalization counterterms
to the low-order diagrams to form a convergent result. In our simplified model, 
only the complex scalar field is quantized and appears within loops in the 
diagrams.

In the covariant formulation, Feynman diagrams are generated
by expanding the effective action for a complex scalar field
\begin{equation}
\mathcal{A}_{\rm eff}=\imu {\rm Tr}\,{\rm log}\left[\partial^2+m^2+V\right]
\label{eq:aeff}\end{equation}
with respect to the components of
\begin{equation}
V=\frac{3m^2}{2}\left(\left|\phi_0\right|^2-1\right)-A_\mu A^\mu 
+\imu\left(A^\mu\overrightarrow{\partial}_\mu
+\overleftarrow{\partial}_\mu A^\mu\right)\,,
\label{eq:aeff1}\end{equation}
where the arrow indicates the direction of differentiation to be
applied in the functional trace, Eq.~(\ref{eq:aeff}). Here we have 
written the effective  action for a general interaction. For the
vortex configuration we will replace $V$ by $V_B$ defined above.

\begin{figure}[ht]

\setlength{\unitlength}{1.6mm}
\parbox[l]{3.5cm}{
\begin{fmfgraph*}(20,20)
  \fmfleft{i3,i4,i1,i,i2,i5,i6}
  \fmfright{o}
  \fmf{phantom,tension=3}{i,v1}
  \fmf{phantom,tension=3}{v2,o}
  \fmf{plain,left=1}{v1,v2,v1}
  \fmf{photon}{v1,i1}
  \fmf{photon}{v1,i2}
\end{fmfgraph*}
}
\parbox[l]{5cm}{
\begin{fmfgraph*}(20,20)
  \fmfleft{i}
  \fmfright{o}
  \fmf{phantom,tension=3}{i,v1}
  \fmf{phantom,tension=3}{v2,o}
  \fmf{plain,left=1}{v1,v2,v1}
  \fmf{photon}{v1,i}
  \fmf{photon}{v2,o}
\end{fmfgraph*}
}
\parbox[l]{4cm}{
\begin{fmfgraph*}(20,20)
  \fmfleft{i}
  \fmfright{o}
  \fmf{phantom,tension=3.5}{i,v1}
  \fmf{phantom,tension=3.5}{v2,o}
  \fmf{plain,left=1}{v1,v2,v1}
  \fmf{plain}{v1,i}
\end{fmfgraph*}
}
\parbox[l]{4cm}{
\begin{fmfgraph*}(20,20)
  \fmfleft{i}
  \fmfright{o}
  \fmf{phantom,tension=4.0}{i,v1}
  \fmf{phantom,tension=4.0}{v2,o}
  \fmf{plain,left=1}{v1,v2,v1}
  \fmf{photon}{v1,i}
\end{fmfgraph*}
}
\vspace{-0.5cm}

\leftline{~\hspace{1.8cm}
a)\hspace{4.2cm}b)\hspace{4.2cm}c) \hspace{3.7cm}d)}
\caption{\label{fig:Qdiv}Quadratically divergent one-loop diagrams.}
\end{figure}
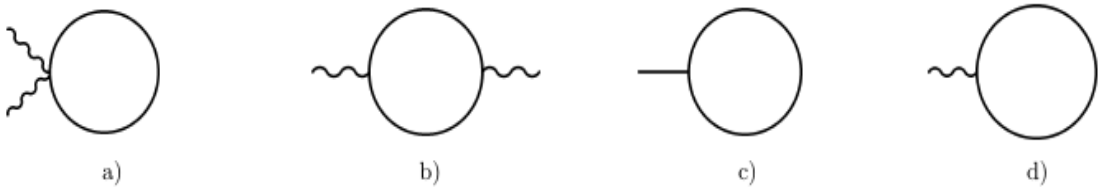

\begin{figure}[ht]

\setlength{\unitlength}{1.5mm}
\parbox[l]{3.5cm}{
\begin{fmfgraph*}(20,20)
  \fmfleft{i3,i4,i1,i,i2,i5,i6}
  \fmfright{o3,o4,o1,o,o2,o5,o6}
  \fmf{phantom,tension=3}{i,v1}
  \fmf{phantom,tension=3}{v2,o}
  \fmf{plain,left=1}{v1,v2,v1}
  \fmf{photon}{v1,i1}
  \fmf{photon}{v1,i2}
  \fmf{photon}{v2,o1}
  \fmf{photon}{v2,o2}
\end{fmfgraph*}
}
\parbox[l]{3.5cm}{
\begin{fmfgraph*}(20,20)
  \fmfleft{i}
  \fmfright{o3,o4,o1,o,o2,o5,o6}
  \fmf{phantom,tension=3}{i,v1}
  \fmf{phantom,tension=3}{v2,o}
  \fmf{plain,left=1}{v1,v2,v1}
  \fmf{photon}{v1,i}
  \fmf{photon}{v2,o1}
  \fmf{photon}{v2,o2}
\end{fmfgraph*}
}
\parbox[l]{3.5cm}{
\begin{fmfgraph*}(20,20)
  \fmfleft{i}
  \fmfright{o3,o4,o1,o,o2,o5,o6}
  \fmf{phantom,tension=1.6}{i,v1}
  \fmf{phantom,tension=1.6}{v2,o3}
  \fmf{phantom,tension=1.6}{v3,o6}
  \fmfpoly{smooth,filled=empty,pull=2.0}{v1,v2,v3}
  \fmf{photon}{v1,i}
  \fmf{photon}{v2,o4}
  \fmf{photon}{v3,o5}
\end{fmfgraph*}
}
\parbox[l]{3.5cm}{
\begin{fmfgraph*}(20,20)
  \fmfleft{i3,i4,i1,i,i2,i5,i6}
  \fmfright{o3,o4,o1,o,o2,o5,o6}
  \fmf{phantom,tension=1.6}{i,v1}
  \fmf{phantom,tension=1.6}{v2,o3}
  \fmf{phantom,tension=1.6}{v3,o6}
  \fmfpoly{smooth,filled=empty,pull=2.0}{v1,v2,v3}
  \fmf{photon}{v1,i1}
  \fmf{photon}{v1,i2}
  \fmf{photon}{v2,o4}
  \fmf{photon}{v3,o5}
\end{fmfgraph*}
}
\parbox[l]{3.5cm}{
\begin{fmfgraph*}(20,20)
  \fmfleft{i3,i4,i1,i,i2,i5,i6}
  \fmfright{o3,o4,o1,o,o2,o5,o6}
  \fmf{phantom,tension=0.8}{v1,i5}
  \fmf{phantom,tension=0.8}{v2,i4}
  \fmf{phantom,tension=0.8}{v3,o4}
  \fmf{phantom,tension=0.8}{v4,o5}
  \fmfpoly{smooth,filled=empty,pull=1.4}{v1,v2,v3,v4}
  \fmf{photon}{v1,i5}
  \fmf{photon}{v2,i4}
  \fmf{photon}{v3,o4}
  \fmf{photon}{v4,o5}
\end{fmfgraph*}
}

\vspace{-0.5cm}

\leftline{~\hspace{1.4cm}
a)\hspace{3.4cm}b)\hspace{3.4cm}c)\hspace{3.4cm}d)\hspace{3.3cm}e)}
\caption{\label{fig:LdivA}Logarithmically divergent one-loop diagrams with 
external photon lines only.}
\end{figure}
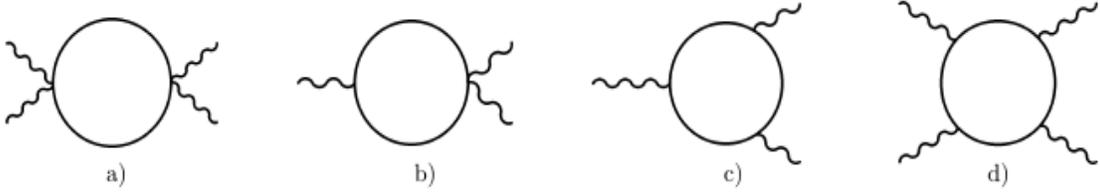

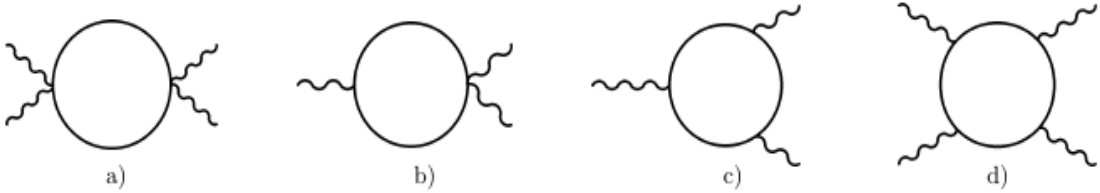
\begin{figure}[ht]

\setlength{\unitlength}{1.5mm}
\parbox[l]{4cm}{
\begin{fmfgraph*}(20,20)
  \fmfleft{i3,i4,i1,i,i2,i5,i6}
  \fmfright{o3,o4,o1,o,o2,o5,o6}
  \fmf{phantom,tension=3}{i,v1}
  \fmf{phantom,tension=3}{v2,o}
  \fmf{plain,left=1}{v1,v2,v1}
  \fmf{photon}{v1,i1}
  \fmf{photon}{v1,i2}
  \fmf{plain}{v2,o}
\end{fmfgraph*}
}
\parbox[l]{4cm}{
\begin{fmfgraph*}(20,20)
  \fmfleft{i}
  \fmfright{o}
  \fmf{phantom,tension=3}{i,v1}
  \fmf{phantom,tension=3}{v2,o}
  \fmf{plain,left=1}{v1,v2,v1}
  \fmf{photon}{v1,i}
  \fmf{plain}{v2,o}
\end{fmfgraph*}
}
\parbox[l]{4cm}{
\begin{fmfgraph*}(20,20)
  \fmfleft{i3,i4,i1,i,i2,i5,i6}
  \fmfright{o}
  \fmf{phantom,tension=2.0}{i1,v1}
  \fmf{phantom,tension=2.0}{i2,v2}
  \fmf{phantom,tension=6.0}{v3,o}
  \fmfpoly{smooth,filled=empty,pull=2.0}{v1,v2,v3}
  \fmf{photon}{i5,v1}
  \fmf{photon}{i4,v2}
  \fmf{plain}{v3,o}
\end{fmfgraph*}
}
\parbox[l]{4cm}{
\begin{fmfgraph*}(20,20)
  \fmfleft{i}
  \fmfright{o}
  \fmf{phantom,tension=3}{i,v1}
  \fmf{phantom,tension=3}{v2,o}
  \fmf{plain,left=1}{v1,v2,v1}
  \fmf{plain}{v1,i}
  \fmf{plain}{v2,o}
\end{fmfgraph*}
}

\vspace{-0.5cm}

\leftline{~\hspace{2.5cm}
a)\hspace{3.8cm}b)\hspace{3.8cm}c)\hspace{3.8cm}d)}
\caption{\label{fig:LdivB}Logarithmically divergent one-loop diagrams with
at least one insertion of the Higgs potential $V_H=\frac{3m^2}{2}(f_H^2-1)$.}
\end{figure}

In Figs.\ \ref{fig:Qdiv}, \ref{fig:LdivA} and \ref{fig:LdivB} we display all
one-loop diagrams that are divergent by power counting. There are considerable
simplifications arising from relationships between the divergent parts of 
different diagrams, but not all of them can be straightforwardly
implemented in Eq.~(\ref{eq:vpe1}). First we note that the divergences from
diagrams with an odd number of gauge fields, \ref{fig:Qdiv}d),
\ref{fig:LdivA}b), \ref{fig:LdivA}c) and \ref{fig:LdivB}b) will have
coefficients that are spatial integrals involving $\partial_\mu A^\mu=0$ for 
the vortex, which will vanish. Hence we do not need to consider these diagrams 
any further. Actually diagrams \ref{fig:Qdiv}d) and \ref{fig:LdivA}b) are 
identically zero for the vortex.

Figure \ref{fig:Qdiv} shows the Feynman diagrams that superficially are 
quadratically divergent. Due to gauge invariance, the
quadratic divergences of a) and b) cancel, so that the only
remaining quadratic divergence is the tadpole graph with a single
insertion of $V_H$ in c). This diagram  is local, {\it i.e.}
independent of the incoming momentum, and  proportional to $\int d^2x
V_H$. Therefore it can be fully removed from the  VPE by an
appropriate {\it no-tadpole} renormalization
condition.

Again by gauge invariance, the logarithmic divergences in
\ref{fig:LdivA}a), \ref{fig:LdivA}d) and \ref{fig:LdivA}e) cancel, as do 
those of \ref{fig:LdivB}a) and \ref{fig:LdivB}c).  Thus all we need to 
subtract in Eq.~(\ref{eq:vpe1}) are the divergences associated with the 
diagrams of Figs.\ \ref{fig:Qdiv}a)-c) and \ref{fig:LdivB}d).  The
treatment of \ref{fig:Qdiv}c) and \ref{fig:LdivB}d) is straightforward
using spectral methods.  We merely need to subtract the Jost functions
obtained from iterating $V_H$ in the differential equations
\begin{equation}
\frac{1}{\rho}\partial_\rho \rho\partial_\rho\overline{\eta}^{(1)}_\ell=
2tZ_\ell(t\rho)\partial_\rho \overline{\eta}^{(1)}_\ell+V_H
\qquad {\rm and}\qquad
\frac{1}{\rho}\partial_\rho \rho\partial_\rho\overline{\eta}^{(2)}_\ell=
2tZ_\ell(t\rho)\partial_\rho \overline{\eta}^{(2)}_\ell
+V_H\overline{\eta}^{(1)}_\ell\,.
\label{eq:BornH1}\end{equation}
Integrating these differential equations with the boundary conditions 
$\lim_{\rho\to\infty} \overline{\eta}^{(1,2)}_\ell(\rho)=0$ 
and $\lim_{\rho\to\infty} \frac{d}{d\rho}
\overline{\eta}^{(1,2)}_\ell(\rho)=0$,
we obtain the subtracted result
\begin{equation}
\left[\nu_\ell(t)\right]_{\ref{fig:Qdiv}c),\ref{fig:LdivB}d)}
=\lim_{\rho\to0}\left\{\overline{\eta}_\ell
-\overline{\eta}^{(1)}_\ell-\overline{\eta}^{(2)}_\ell
+\frac{1}{2}\left(\overline{\eta}^{(1)}_\ell\right)^2\right\}\,.
\label{eq:BornH2}\end{equation}

Next we explain how to complete a full set of subtractions corresponding to 
the remaining divergences. However, in the next section we will show a 
shortcut that will bypass these subtractions in favor of a simpler limiting 
function, which we will then use in the context of the singular vortex 
background in Sections \ref{sec:SB} and \ref{sec:NE}. Diagram \ref{fig:Qdiv}a) 
has a single insertion of $(f_G/\rho)^2$ and the corresponding
order of the Jost solution is given by the differential equation
\begin{equation}
\frac{1}{\rho}\partial_\rho \rho\partial_\rho\overline{\eta}^{(3)}_\ell
=2tZ_\ell(t\rho)\partial_\rho \overline{\eta}^{(3)}_\ell
+\left(\frac{f_G}{\rho}\right)^2\,.
\label{eq:BornH3}\end{equation}
To obtain the Jost solution representing diagram \ref{fig:Qdiv}b), which 
has two distinct insertions of $2\ell f_G/\rho^2$, we need to
solve a set of coupled differential equations
\begin{equation}
\frac{1}{\rho}\partial_\rho \rho\partial_\rho\overline{\eta}^{(4)}_\ell
=2tZ_\ell(t\rho)\partial_\rho \overline{\eta}^{(4)}_\ell
-\frac{2\ell}{\rho^2}f_G
\qquad {\rm and}\qquad
\frac{1}{\rho}\partial_\rho \rho\partial_\rho\overline{\eta}^{(5)}_\ell
=2tZ_\ell(t\rho)\partial_\rho \overline{\eta}^{(5)}_\ell
-\frac{2\ell}{\rho^2}f_G\overline{\eta}^{(4)}_\ell\,.
\label{eq:BornH4}\end{equation}
Again, these differential equations are solved subject to the boundary
conditions $\lim_{\rho\to\infty}\overline{\eta}^{(3,4,5)}_\ell(\rho)=0$ and
$\lim_{\rho\to\infty} \frac{d}{d\rho}
\overline{\eta}^{(3,4,5)}_\ell(\rho)=0$. Putting things together, 
the integral in Eq.~(\ref{eq:vpe1}) is rendered finite by using
\begin{equation}
\left[\nu_\ell(t)\right]_n=\lim_{\rho\to0}\left\{\overline{\eta}_\ell
-\overline{\eta}^{(1)}_\ell-\overline{\eta}^{(2)}_\ell
+\frac{1}{2}\left(\overline{\eta}^{(1)}_\ell\right)^2
-\overline{\eta}^{(3)}_\ell-\overline{\eta}^{(4)}_\ell
-\overline{\eta}^{(5)}_\ell
+\frac{1}{2}\left(\overline{\eta}^{(4)}_\ell\right)^2\right\}\,.
\label{eq:BornH5}\end{equation}
Note that we have included $\overline{\eta}^{(4)}_\ell$ in the subtraction 
even though $\sum_{\ell=-L}^{L}\overline{\eta}^{(4)}_\ell=0$
because it may be advantageous numerically by avoiding summing large numbers.

Before discussing results, a remark on the numerical
simulation is in order.  Scattering calculations with radial symmetry
separate the regular and irregular solutions at the origin. In the
$\ell=0$ channel, the irregular solution diverges like a
logarithm, while the regular one is constant. Numerically these two
behaviors are difficult to disentangle. We therefore integrate the
differential equations for $\ell=0$ down to various values in the
regime $\rho_{\rm min}=10^{-20}\ldots 10^{-50}/m$ and fit
$c_0+c_1/\ln[\rho_{\rm min}/m]+c_2/\ln^2[\rho_{\rm min}/m]$. 
We then make sure that the final result, $c_0$, is stable
under further variation of $\rho_{\rm min}$.  Fortunately, we will
see later that an appropriate regularization can eliminate the need
for this detailed analysis.

\begin{figure}
\centerline{
\epsfig{file=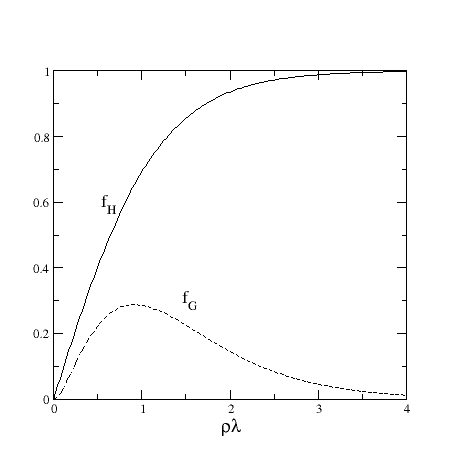,width=4cm,height=5cm}\hspace{1cm}
\epsfig{file=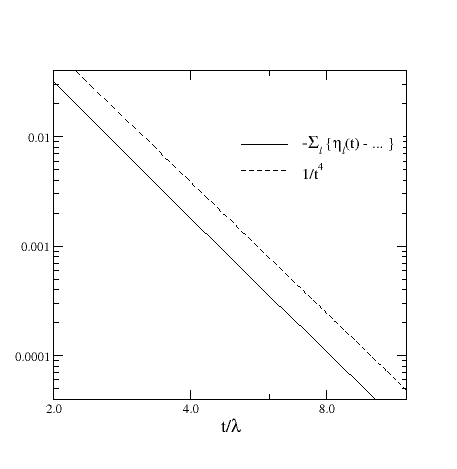,width=9cm,height=5cm}}
\caption{\label{fig:reg1}Left panel: regular vortex background profile functions;
right panel: double logarithmic plot of $\sum_\ell\left[\eta_\ell(t)\right]_n$
from this background  compared to a power behavior, both with the
horizontal axis expressed in units of $\lambda = m/\sqrt{2}$.}
\end{figure}

In Fig.~\ref{fig:reg1} we display both a sample regular vortex background and
the resulting integrand $\sum_\ell\left[\eta_\ell(t)\right]_n$ for the VPE.
Obviously, that integrand quickly approaches zero like $1/t^4$, which suffices 
to make the integral in Eq.~(\ref{eq:vpe1}) converge. Hence we have correctly 
implemented the divergence structure deduced from the Feynman diagrams in the
scattering  calculation.

Surprisingly
\begin{equation}
\lim_{L\to\infty}\sum_{\ell=}^{L}\, \lim_{\rho\to0}\left\{
\overline{\eta}^{(3)}_\ell+\overline{\eta}^{(4)}_\ell+
\overline{\eta}^{(5)}_\ell
-\frac{1}{2}\left(\overline{\eta}^{(4)}_\ell\right)^2\right\}
\,\longrightarrow\, \int_0^\infty \frac{d\rho}{\rho} f_G^2(\rho)
\qquad {\rm as}\qquad t\,\to\,\infty\,,
\label{eq:LinInt}\end{equation}
signaling a quadratic divergence. The integral is the coefficient of the quadratic 
divergence arising when the regularization procedure is not gauge invariant, which here 
is a consequence of the fact that the Jost function is not gauge invariant. We will
describe how to fix this problem in the discussion below.

\end{fmffile}

\section{Fourier transforms and Feynman diagrams}
\label{sec:FT}

As far as the gauge field part of the background is concerned, we have
argued above  that even though there are many divergent diagrams, only 
the divergences from \ref{fig:Qdiv}a) and b) will remain once all 
divergent diagrams are combined with a gauge invariant regularization.

Let us first comment on the origin of Eq.~(\ref{eq:LinInt}). In a
covariant, but not necessarily gauge invariant regularization,
diagrams \ref{fig:Qdiv}a) and b) contribute 
\begin{equation}
\int d^4x  A_\mu(x)A_\nu(x) \int \frac{d^4l}{(2\pi)^4}\,
\left[\frac{g^{\mu\nu}}{l^2-m^2+\imu0^+}
-\frac{1}{2}\frac{4l^\mu l^\nu}{(l^2-m^2+\imu0^+)^2}\right]
\label{eq:QD1}\end{equation}
as the leading divergence in the effective action. In dimensional regularization 
the two terms cancel. However, using a sharp cut-off $\Lambda$ in Euclidean space
yields
\begin{equation}
\frac{\Omega_D}{D}\frac{\Lambda^D}{\Lambda^2+m^2}
-\Omega_D\int_0^\Lambda dl
\left[\frac{l^D}{D}\frac{d}{dl}\left(\frac{1}{l^2+m^2}\right)
+\frac{2}{D}\frac{l^{D+1}}{(l^2+m^2)^2}\right]
\label{eq:cutoff1}\end{equation}
times $g^{\mu\nu}$ for the loop momentum integral, where $\Omega_D$ is
the $D$-dimensional solid angle. The integrand indeed vanishes for any
$D$, but the surface term does not vanish for $D\ge2$ as the regulator
is removed.\footnote{Dimensional regularization takes $D$ small
enough such that surface term tends to zero as $\Lambda\to\infty$.} With 
$A_\mu(x)A^\mu(x)=-\left(\frac{f_G}{\rho}\right)^2$ we observe that
the quadratic divergence for $D=2$ is proportional to the coefficient
given in Eq.~(\ref{eq:LinInt}).

In dimensional regularization (with $D$ space time dimensions)
the diagrams \ref{fig:Qdiv}a) and b) contribute
\begin{equation}
E_{\rm VPE}^{(A)}=\Gamma\left(1-\frac{D}{2}\right)
\left(\frac{4\pi\mu^2}{m^2}\right)^{2-D/2}\left(\frac{m}{4\pi}\right)^2
\int\frac{d^2p}{(2\pi)^2}\,
\widetilde{\Vek{A}}(\Vek{p})\cdot\widetilde{\Vek{A}}(\Vek{-p})
\int_0^1 dx\,
\left\{1-\left[1+x(1-x)\frac{\Vek{p}^2}{m^2}\right]^{D/2-1}\right\}
\label{eq:FDA}\end{equation}
to the VPE of the vortex per unit length. The derivation of
Eq.~(\ref{eq:FDA}) used that $\partial_\mu A^\mu=0$
translates into $\Vek{p}\cdot\widetilde{\Vek{A}}(\Vek{p})=0$. 
The renormalization scale $\mu$ will eventually be eliminated by appropriate 
renormalization conditions. The pole at $D=2$, corresponding to the
superficial quadratic divergence, has zero residue and we may thus
analytically continue to $D=4-2\epsilon$ without adding a counterterm
for this divergence. The logarithmic divergence in
Eq.~(\ref{eq:FDA}) becomes
\begin{equation}
E_{\rm VPE}^{(A)}\Big|_{\rm div.}=\frac{1}{6\epsilon(4\pi)^2}
\int\frac{d^2p}{(2\pi)^2}\, \Vek{p}^2 
\widetilde{\Vek{A}}(\Vek{p})\cdot\widetilde{\Vek{A}}(\Vek{-p})
=\frac{1}{12\epsilon(4\pi)^2}\int d^2x\, F_{\mu\nu}F^{\mu\nu}\,.
\label{eq:FDA1}\end{equation}
Using the inverse Fourier transformation to write a coordinate space integral
is actually essential. As we will see later, momentum space integrals
above may not be well-defined for singular vector potential, while the
field strength tensor only contains the magnetic field, which is
non-singular. In a full theory the ultra-violet divergence in
Eq.~(\ref{eq:FDA1}) is compensated by wave-function renormalization
of the dynamical gauge field. We therefore refer to this divergence
as wave-function renormalization.

In the next step we translate the logarithmic divergence of
Eq.~(\ref{eq:FDA1}) into the large $t$-behavior of $\nu(t)$. We write,
in terms of an arbitrary mass scale $M$,
\begin{equation}
\frac{1}{\epsilon(4\pi)^2}=-\imu\int\frac{d^4l}{(2\pi)^4}\,
\frac{1}{\left(l^2-M^2+\imu0^+\right)^2}\Bigg|_{\rm div.}
=\frac{1}{8\pi^2}\int \frac{l^2 dl}{\sqrt{l^2+M^2}^3}\Bigg|_{\rm div.}\,,
\label{eq:lf1}\end{equation}
so that we may formally identify
\begin{equation}E_{\rm VPE}^{(A)}\Bigg|_{\rm div.}=
\frac{1}{96\pi^2}\left[\int d^2x\, F_{\mu\nu}F^{\mu\nu}\right]
\int \frac{l^2 dl}{\sqrt{l^2+M^2}^3}\Bigg|_{\rm div.}\,.
\label{eq:lf2}\end{equation}
Hence with only gauge field insertions ({\it i.e.} $V_H=0$),
the un-subtracted integrand in Eq.~(\ref{eq:vpe1}) has the asymptotic
behavior
\begin{equation}
\left[\nu(t)\right]_0 - \int_0^\infty \frac{d\rho}{\rho} f_G^2(\rho)\,
\longrightarrow\,\nu_{\rm l.f.}(t)
=\frac{1}{t^2}\,\frac{1}{12}\int_0^\infty \rho d\rho
\left(\frac{f_G^\prime(\rho)}{\rho}\right)^2
\qquad {\rm as}\qquad t\,\longrightarrow\,\infty\,,
\label{eq:lf3}\end{equation}
where $F_{\mu\nu}F^{\mu\nu}=2\left(\frac{f_G^\prime(\rho)}{\rho}\right)^2$
for the  vortex. With $V_H \neq 0$,
the limiting function should also be approached when the
divergences arising from $V_H$ are removed, {\it i.e.} by
using Eq.~(\ref{eq:BornH2}) instead of Eq.~(\ref{eq:BornH5}) and
subtracting the constant of Eq.~(\ref{eq:LinInt}).

\begin{figure}[t]
\centerline{
\epsfig{file=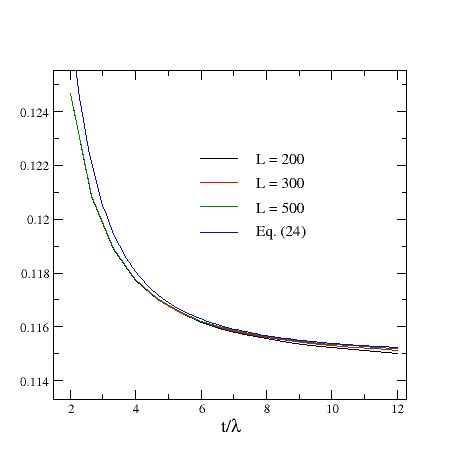,width=8cm,height=5cm}\hspace{1cm}
\epsfig{file=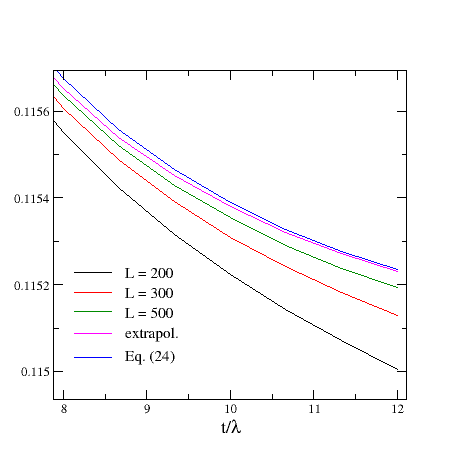,width=8cm,height=5cm}}
\caption{\label{fig:asym1}(Color online) Comparison of Born approximation
and asymptotics extracted from Feynman diagrams, both with the
horizontal axis expressed in units of $\lambda = m/\sqrt{2}$.}
\end{figure}

We can restate the above analysis of the ultra-violet behavior by 
supplementing Eq.~(\ref{eq:LinInt}) with the next to leading order term
\begin{equation}
\lim_{L\to\infty}\sum_{\ell=-L}^{L}\, \lim_{\rho\to0}\left\{
\overline{\eta}^{(3)}_\ell+\overline{\eta}^{(4)}_\ell+
\overline{\eta}^{(5)}_\ell
-\frac{1}{2}\left(\overline{\eta}^{(4)}_\ell\right)^2\right\}
\,\longrightarrow\, \int_0^\infty \frac{d\rho}{\rho} f_G^2(\rho)
+\frac{1}{12t^2}\,\int_0^\infty \rho d\rho
\left(\frac{f_G^\prime(\rho)}{\rho}\right)^2
\label{eq:LinIntB}\end{equation}
as $t\,\to\,\infty$. We numerically calculate this asymptotic expression in 
Fig.~\ref{fig:asym1}. For this calculation we need to terminate the sum on 
the left hand side at some large but finite value.
From the left panel of this figure we see that the leading terms of the 
Born series nicely approach the expected asymptotic form for moderate
momenta $t$. However, closer inspection (right panel) shows that even for 
values as large as $L=500$, the asymptotic form is not exactly reached.
Rather, the difference between the sum and the asymptotic form increases 
with $t$. For this reason we use various (large) values for $L$ and fit 
an inverse polynomial
\begin{equation}
\sum_{\ell=-L}^{L}\,
\lim_{\rho\to0}\left\{
\overline{\eta}^{(3)}_\ell+\overline{\eta}^{(4)}_\ell+
\overline{\eta}^{(5)}_\ell
-\frac{1}{2}\left(\overline{\eta}^{(4)}_\ell\right)^2\right\}
=c_0+\frac{c_1}{L}+\frac{c_2}{L^2}+\ldots
\label{eq:fit1}\end{equation}
so that $c_0$ is the $L\to\infty$ extrapolation. Indeed, we see from the 
right panel of Fig.~\ref{fig:asym1} that this extrapolation perfectly matches 
the asymptotic form extracted from the analysis of the Feynman diagrams. This 
result shows that we do not miss any subleading logarithmic divergences when
we subtract the constant in Eq.~(\ref{eq:LinInt}). For regular profiles, the 
numerical effect of this extrapolation is actually very marginal. However, we 
will see later that for singular vortex profiles this extrapolation is essential 
and its omission leads to incorrect results.

In further preparation for the consideration of
singular gauge profiles, it is illuminating to consider the finite
part in Eq.~(\ref{eq:FDA}). In $\overline{\rm MS}$
renormalization, it reads
\begin{align}
\overline{E}_{\rm VPE}^{(A)}&=
-\left(\frac{m}{4\pi}\right)^2\int\frac{d^2p}{(2\pi)^2}\,
\widetilde{\Vek{A}}(\Vek{p})\cdot\widetilde{\Vek{A}}(-\Vek{p})
\int_0^1dx\,\left[1+x(1-x)\frac{\Vek{p}^2}{m^2}\right]
\ln\left[1+x(1-x)\frac{\Vek{p}^2}{m^2}\right]\cr
&=-\frac{m^2}{8\pi}\int_0^\infty pdp\, a^2(p)
\int_0^1dx\,\left[1+x(1-x)\frac{p^2}{m^2}\right]
\ln\left[1+x(1-x)\frac{p^2}{m^2}\right]\,,
\label{eq:EAmsbar}\end{align}
where $a(p)=\int_0^\infty d\rho\, f_G(\rho)J_1(\rho p)$ (which indeed exists 
even for a singular background).
Straightforward substitution of Eq.~(\ref{eq:FT1}) fails for profiles with
$f_G(0)\ne0$, as can be seen already from the leading $\Vek{p}^2$ term.  
Integration by parts of $\frac{d}{d\rho}J_0(p\rho)=-pJ_1(p\rho)$ and the 
completeness relation for the Bessel functions yield
\begin{equation}
\int_0^\infty pdp\, p^2 a^2(p)=
f_G^2(0)\int_0^\infty pdp+2f_G(0)\int_0^\infty pdp
\left[\int_0^\infty d\rho\, J_0(p\rho)f_G^\prime(\rho)\right]
+\int_0^\infty \rho d\rho\,\left(\frac{f_G^\prime(\rho)}{\rho}\right)^2\,.
\label{eq:failFT}\end{equation}
The first two integrals diverge, while the last term is the
correct local integral $\int \frac{d^2x}{4\pi}
F_{\mu\nu}F^{\mu\nu}$. In order to compute Eq.~(\ref{eq:EAmsbar})
for a singular gauge profile, we first note that the Feynman
parameter integral in  Eq.~(\ref{eq:EAmsbar}) vanishes for
$\Vek{p}^2=0$, so that
\begin{equation}
\overline{E}_{\rm VPE}^{(A)}=
-\left(\frac{1}{4\pi}\right)^2\int\frac{d^2p}{(2\pi)^2}\,\Vek{p}^2
\widetilde{\Vek{A}}(\Vek{p})\cdot\widetilde{\Vek{A}}(-\Vek{p})
\int_0^1dx\,\left[\frac{m^2}{\Vek{p}^2}+x(1-x)\right]
\ln\left[1+x(1-x)\frac{\Vek{p}^2}{m^2}\right]
\label{eq:EAmsbar1}\end{equation}
is well defined. Next we observe that (formally)
\begin{equation}
\int\frac{d^4p}{(2\pi)^4}\, p_\mu \widetilde{A}_\nu(p) p^\mu
\widetilde{A}^\nu(-p)g(p)=\frac{1}{2}\int\frac{d^4p}{(2\pi)^4}\,
\widetilde{F}_{\mu\nu}(p)\widetilde{F}^{\mu\nu}(-p)g(p)\,,
\label{eq:FT2}\end{equation}
again using $\partial_\mu A^\mu=0$. We then define
\begin{equation}
b(p)=-\int_0^\infty d\rho\, f_G^\prime(\rho) J_0(p\rho)
\label{eq:FT3}\end{equation}
to find
\begin{equation}
\overline{E}_{\rm VPE}^{(A)}= -\frac{1}{8\pi}\int_0^\infty pdp\,
b^2(p)\int_0^1dx\,\left[\frac{m^2}{p^2}+x(1-x)\right]
\ln\left[1+x(1-x)\frac{p^2}{m^2}\right]\,.
\label{eq:EAmsbar2}\end{equation}
For regular profiles we have numerically verified that Eqs.~(\ref{eq:EAmsbar})
and~(\ref{eq:EAmsbar2}) yield identical results. However, Eq.~(\ref{eq:EAmsbar2})
is also well-defined for singular vortex profiles.

\section{Singular background}
\label{sec:SB}

The standard procedure outlined in Sec. \ref{sec:standard} fails when
$f_G(0)\ne0$.  The situation is even worse than Eq.~(\ref{eq:failFT})
suggests because neither the Born approximation nor the Fourier
transform of the radial function $\left[f_G^2-2\ell f_G\right]/\rho^2$
exist. Yet we expect
\begin{equation}
\left[\nu(t)\right]_{H}=\lim_{L\to\infty}\sum_{\ell=-L}^{L}
\left\{\overline{\eta}_\ell
-\overline{\eta}^{(1)}_\ell-\overline{\eta}^{(2)}_\ell
+\frac{1}{2}\left(\overline{\eta}^{(1)}_\ell\right)^2\right\}
\Bigg|_{\rho=\rho_{\rm min}}
-\int_{\rho_{\rm min}}^\infty \frac{d\rho}{\rho}\, f_G^2(\rho)
\label{eq:sing1}\end{equation}
to approach $\nu_{\rm l.t.}(t)$ as $\rho_{\rm min}\to0$ and
$t\to\infty$, to properly 
produce the logarithmic divergence. Essentially we have subtracted the
ultra-violet divergences associated with the scalar potential as in
Eq.~(\ref{eq:BornH2}) and subtracted the constant, which formally does
not contribute to the VPE, needed to maintain the analytical properties 
of the (summed logarithm of the) Jost function. The above analysis refers 
to the scattering data for imaginary momenta, but we have also computed 
the Jost function for real momenta. In that case the subtracted constant 
emerges as the modulus of the Jost function and thus does not contribute 
to the VPE which, on the real axis, involves only the phase of the Jost
function and the bound state energies.

We have numerically verified that the regulator $\rho_{\rm min}$ in
that integral must be identical to the end-point of differential
equation. Surprisingly, we also find numerically that, with this
regularization installed, the sensitivity on $\rho_{\rm min}$ is
mitigated  and the extrapolation procedure of Sec. \ref{sec:standard}
is no longer necessary.

To construct the limiting function that will allow us to implement the final 
logarithmic renormalization, we require a background potential that has the 
same logarithmic divergence in its second  order Born term. For this purpose 
we introduce $V_f=\frac{3m^2}{2}(\tanh^2(\rho)-1)$, which couples to the 
scalar quantum fluctuations via Feynman diagrams like \ref{fig:Qdiv}c) and 
\ref{fig:LdivB}d). All we require are the solutions to Eq.~(\ref{eq:BornH1}) 
with $V_H$ replaced by $V_f$. Calling those solutions 
$\overline{\zeta}_\ell^{(1,2)}(\rho)$, we find the second order contribution
\begin{equation}
\overline{\nu}^{(2)}(t)=\lim_{L\to\infty}\sum_{\ell=0}^{L}
\left(2-\delta_{\ell0}\right)\lim_{\rho\to0}
\left[\overline{\zeta}_\ell^{(2)}(\rho)
-\frac{1}{2}\left(\overline{\zeta}_\ell^{(1)}(\rho)\right)^2\right]\,.
\label{eq:lf4}\end{equation}
Similarly to Eq.~(\ref{eq:lf3}), it produces the limiting function
\begin{equation}
\overline{\nu}^{(2)}(t)\,\longrightarrow\,\overline{\nu}_{\rm l.f.}(t)
=-\frac{1}{4t^2}\,\int_0^\infty \rho d\rho\, V_f^2 
\qquad {\rm as}\qquad t\,\longrightarrow\,\infty\,.
\label{eq:lf5}\end{equation}
Using this quantity we compute the finite scattering data component of
the VPE as
\begin{equation}
E_{\rm VPE}^{\rm scat.}=\frac{1}{2\pi}\int_m^\infty tdt\,
\left[\left[\nu(t)\right]_H-c_B \overline{\nu}^{(2)}(t)\right]
\qquad {\rm with}\qquad
c_B=-\frac{1}{3}\frac{\int_0^\infty \rho d\rho
\left(\frac{f_G^\prime(\rho)}{\rho}\right)^2}
{\int_0^\infty \rho d\rho\, V_f^2}\,.
\label{eq:almost_finished}\end{equation}
Instead of the wave-function renormalization,
Eq.~(\ref{eq:EAmsbar2}), we have to add back the Feynman
diagrams for two scalar field insertions, both the regular,
$V_H$, and the auxiliary, $V_f$. In $\overline{\rm MS}$
renormalization they are combined to
\begin{equation}
\overline{E}_{\rm VPE}^{(S)}= \frac{1}{16\pi}\int_0^\infty pdp\,
\left[v_H^2(p)+c_Bv_f^2(p)\right]\int_0^1dx\,
\ln\left[1+x(1-x)\frac{p^2}{m^2}\right]
\quad {\rm where}\quad
v_{H,f}(p)=\int_0^\infty \rho d\rho\, V_{H,f}(\rho)J_0(p\rho)\,.
\label{eq:ESmsbar}\end{equation}
In total, the VPE for the singular background is given by
\begin{equation}
E_{\rm VPE}=E_{\rm VPE}^{\rm scat.}+\overline{E}_{\rm VPE}^{(S)}\,.
\label{eq:master}\end{equation}

\section{Numerical experiments}
\label{sec:NE}

For the numerical simulations we consider a family of (singular)
background profiles
\begin{equation}
f_H(\rho)=\tanh(\alpha\rho\lambda)
\qquad{\rm and}\qquad
f_G(\rho)={\rm e}^{-(\alpha\rho\lambda)^2}
\label{eq:profile}\end{equation}
that have shapes similar to the ANO profiles. Here $\alpha$ is a 
dimensionless variational parameter and $\lambda = m/\sqrt{2}$ provides a 
convenient energy (or, equivalently, inverse length) scale. Though the 
gauge potential is singular at $\rho\to0$, the profile of the magnetic field, 
$\frac{f_G^\prime(\rho)}{\rho}$, has a smooth limit. To study the relative 
roles of the scalar and gauge potentials, we introduce the additional 
parameter $\beta$ and define $V_H=(3m^2/2)\beta\left[f^2_H(\rho)-1\right]$. 
Note that the model Lagrangian, Eq.~(\ref{eq:lag}), implies $\beta=1$. 
However, our main interest is to compare the scattering results based on 
Eq.~(\ref{eq:flct2}) with the canonical expansion of the effective action, 
Eq.~(\ref{eq:aeff}). Then, {\it e.g.} a small value of $\beta$ is an ideal
means to investigate the role of the singularity in $\Vek{A}$.

In order to obtain numerically stable results when integrating 
differential equations (\ref{eq:jost2}), (\ref{eq:BornH1}), etc.\@
we needed to develop a long-double precision {\sf fortran} code for an 
adaptive step size control in combination with a fourth order 
Runge-Kutta algorithm~\cite{NR}. The algorithm for computing the modified 
Bessel function $K_{|\ell|}(t\rho)$ is an iterative process in $\ell$, 
which is computationally costly for large $\ell$. Some of the simulations 
that did not require extremely large wave numbers and angular momenta were 
cross-checked with computations in {\sf mathematica}.

\begin{figure}
\centerline{
\epsfig{file=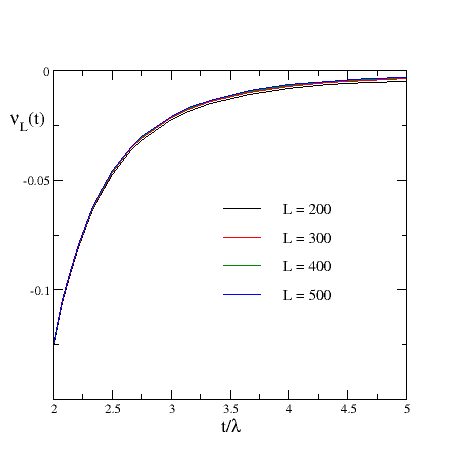,width=5.5cm,height=4cm}\hspace{0.5cm}
\epsfig{file=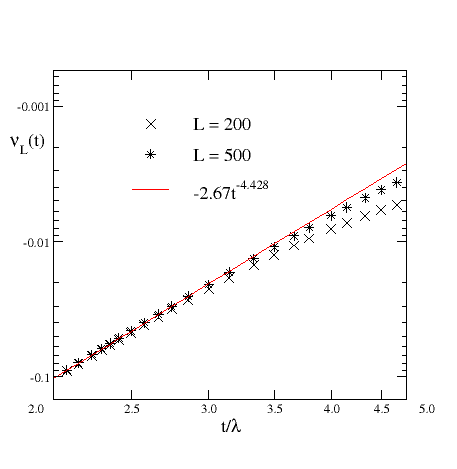,width=5.5cm,height=4cm}\hspace{0.5cm}
\epsfig{file=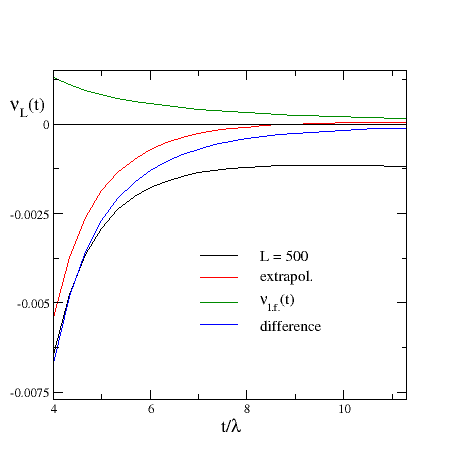,width=5.5cm,height=4cm}}
\caption{\label{fig:num1}(Color online) Asymptotic behavior of
$\nu_L(t)$ defined in Eq.~(\ref{eq:sing2}) using $\alpha=0.5$ and
$\beta=1.0$ in Eq.~(\ref{eq:profile}). Left panel: different maximal 
angular momenta $L$;  center panel: fit showing that $\nu_L(t)$ falls 
faster than $1/t^2$ for moderate values of $t$; right panel: extrapolation 
$L\to\infty$ and limiting function from Eq.~(\ref{eq:lf3}), all with the
horizontal axis expressed in units of $\lambda = m/\sqrt{2}$.  Also
shown is the difference between the extrapolated and limiting functions.}
\end{figure}

\begin{figure}
\centerline{
\epsfig{file=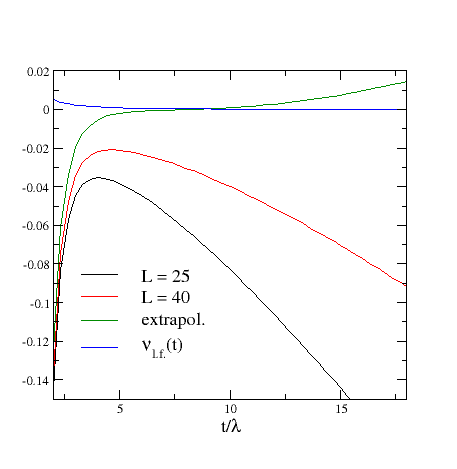,width=5.5cm,height=4.5cm}\hspace{0.5cm}
\epsfig{file=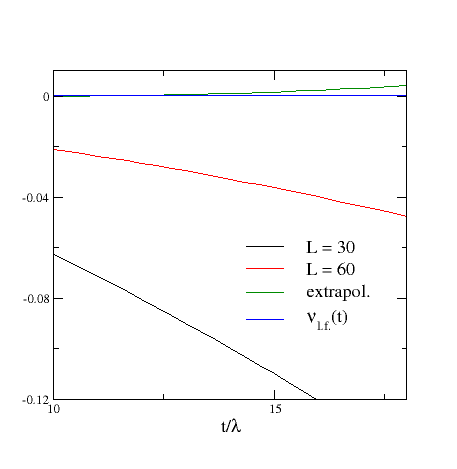,width=5.5cm,height=4.5cm}\hspace{0.5cm}
\epsfig{file=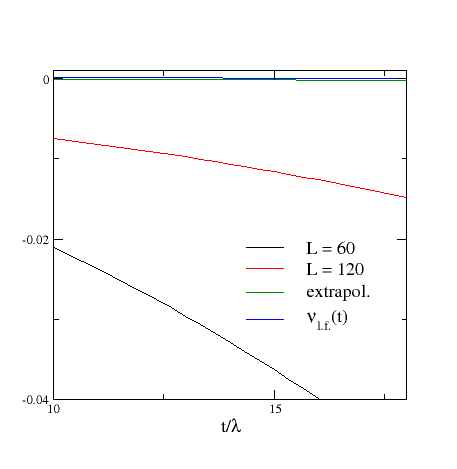,width=5.5cm,height=4.5cm}}
\caption{\label{fig:num2}(Color online) Asymptotic structure of the
angular momentum sum in Eq.~(\ref{eq:sing2}) for $\alpha=0.5$ and $\beta=1.0$. 
Four equi-distant $L$ values between (and including) the explicitly given
ones are used to match the right-hand-side as in
Eq.~(\ref{eq:fit1}). Note the different intervals for the horizontal
axes, which are given in units of $\lambda = m/\sqrt{2}$.}
\end{figure}

To begin with, we have to verify that Eq.~(\ref{eq:sing1}) indeed
approaches the limiting function, Eq~(\ref{eq:lf3}), when the
imaginary momentum $t$ becomes large.  For this purpose it is useful to
introduce a variant of Eq.~(\ref{eq:sing1}) without the $L\to\infty$ limit
\begin{equation}
\nu_L(t)=\sum_{\ell=-L}^L
\left\{\overline{\eta}_\ell
-\overline{\eta}^{(1)}_\ell-\overline{\eta}^{(2)}_\ell
+\frac{1}{2}\left(\overline{\eta}^{(1)}_\ell\right)^2\right\}
\Bigg|_{\rho=\rho_{\rm min}}
-\int_{\rho_{\rm min}}^\infty \frac{d\rho}{\rho}\, f_G^2(\rho)\,.
\label{eq:sing2}\end{equation}
As a first step we have confirmed that the results are indeed
insensitive to the choice of $\rho_{\rm min}$ once it is taken small
enough.  The dependence on $L$ is much more
intricate. A typical result is shown in Fig.~\ref{fig:num1}.  At
first sight (left panel) we observe a reasonably convergent 
function as $L$ becomes very big. The graphs indicate that the sums indeed
approach zero at large $t$.  Yet there are problems with those curves. 
The double-logarithmic plot in the center panel of Fig.~\ref{fig:num2} 
strongly suggests that $\nu_L(t)$ approaches zero much faster than 
$1/t^2$ for moderate $t$. Hence using $\nu_L(t)$ instead 
of $\left[\nu(t)\right]_n$ in Eq.~(\ref{eq:vpe1}) would produce a finite 
result without wave-function renormalization of the gauge field.\footnote{We 
remark that a similar behavior was observed for the fermion VPE of 
a vortex \cite{Pasipoularides:2000gg}.}
Alternatively, we might be missing a sub-leading logarithmic divergence. This, 
however, would invalidate our results for regular profiles. Either way, the
observed behavior of $\nu_L(t)$ looks like a clear contradiction of
renormalizability. Even worse, $\nu_L(t)$ seems to be negative for all
values of $t$, while the limiting function, Eq.~(\ref{eq:lf3}), is positive 
as we observed for regular profiles. It is therefore unavoidable to
investigate the large $t$ and large $L$ behavior in more detail.
The right panel of Fig.~\ref{fig:num1} also shows the result of an
extrapolation for $\nu_L(t)$ as in Eq.~(\ref{eq:fit1}). We observe
that the difference between the extrapolated and the $L=500$ curves 
increases with $t$. Eventually the extrapolated curve indeed crosses 
zero and approaches the limiting function such that the difference 
decays like $1/t^4$.  We observe this behavior also when fitting 
$\nu_L(t)$ for $t/\lambda\in[4,8]$. For $L=500$ we find
$\nu_{500}(t)\approx-0.039\lambda^2/t^2-0.966\lambda^4/t^4\,(0.995)$,
while the  extrapolated function follows
$\nu_{\infty}(t)\approx0.023
\lambda^2/t^2-1.750\lambda^4/t^4\,(1.000)$; the data in
parentheses are the correlation coefficients of the respective
fits. For $t/\lambda\in[8,14]$ we find $\nu_{\infty}(t)\approx 0.021
\lambda^2/t^2-1.640\lambda^4/t^4\,(1.000)$,  while
$\nu_{500}(t)$ departs from zero for $t>10 \lambda$ and we cannot reasonably
approximate it with a polynomial of inverse momenta in this regime.
Recalling that $\nu_{\rm l.f.}(t)\approx0.021 \lambda^2/t^2$ we conclude that
$\nu_{\rm l.f.}(t)$ is not matched for any finite $L$ and that the
extrapolation to infinitely large angular momenta is essential. 

For any fixed value of $L$, there is a critical momentum $t$ above
which $\nu_L(t)$ deviates from the convergence pattern suggested by
the left panel of Fig.~\ref{fig:num1}, where it may even acquire a negative
slope as a function of $t$. In our numerical simulation we were able
to generate stable results for $L\lesssim600$. However, even that
many partial waves was not large enough to obtain any
positive result for $\nu_L(t)$ when using $\alpha=0.5$. Said another way, 
without the extrapolation we get nowhere near the limiting function.  This
extrapolation is thus essential for the singular background, because 
otherwise $\nu_L(t)$ deviates significantly from the limiting function at
moderate momenta (unlike the case of regular backgrounds). This deviation will 
only be overcome at large imaginary momenta $t$, which in turn requires
us to include extremely large angular momenta.

It is worthwhile to ask whether such large $L$ values are needed to
obtain a reliable extrapolation. The results of the corresponding
simulations are shown in Fig.~\ref{fig:num2}. Obviously,
extrapolations that fit the right-hand-side of Eq.~(\ref{eq:fit1}) in
the vicinity of $L=50$ do not match the limiting function and actually
produce asymptotic behavior that departs from zero already at
moderate momenta. An extrapolation based on results from $L\sim100$ 
appears to do better, but a closer inspection also reveals that the 
extrapolation in the right panel in Fig.~\ref{fig:num2} has a maximum 
at $t\approx7 \lambda$ and departs from zero for larger momenta. These 
results cast doubt on the numerical simulations of Ref.~\cite{Baacke:2008sq}, 
which employed $L\in[25,35]$ for the extrapolation.
One may argue that the numerical error that arises from not matching 
the correct asymptotic behavior might be negligible and one could simply 
omit the contribution to the VPE arising from moderate and large $t$. 
Note, however, that there is an additional factor $t$ in the integrand 
of Eq.~(\ref{eq:vpe1}), and that the value of $t$ at which that
integral may eventually be truncated can only be determined 
{\it a posteriori}. The essential criterion is that there is a
significantly large interval along the $t$ axis in which the numerical
simulation produces a $1/t^4$ behavior for the difference between the
extrapolation and the limiting function.

\begin{figure}
\centerline{
\epsfig{file=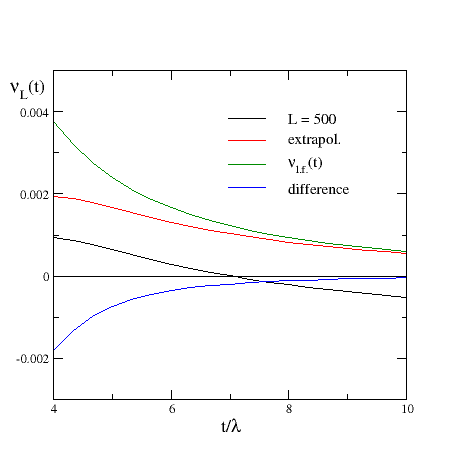,width=7cm,height=4cm}\hspace{1cm}
\epsfig{file=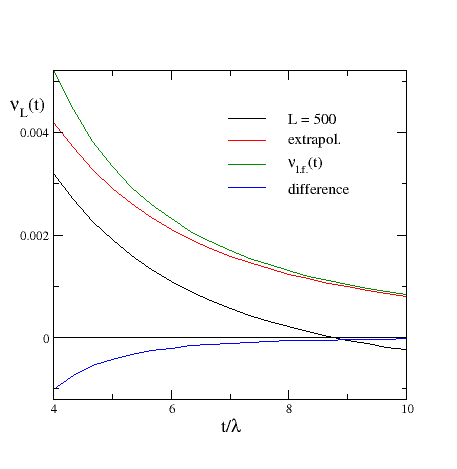,width=7cm,height=4cm}}
\centerline{
\epsfig{file=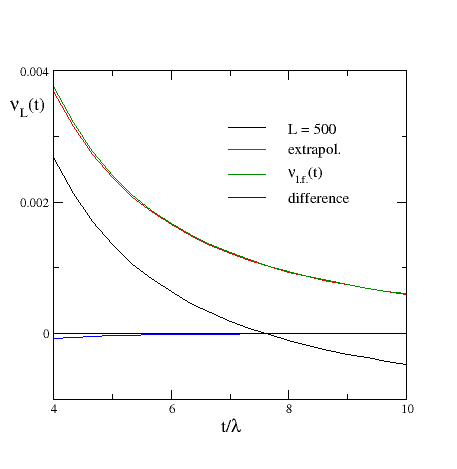,width=7cm,height=4cm}\hspace{1cm}
\epsfig{file=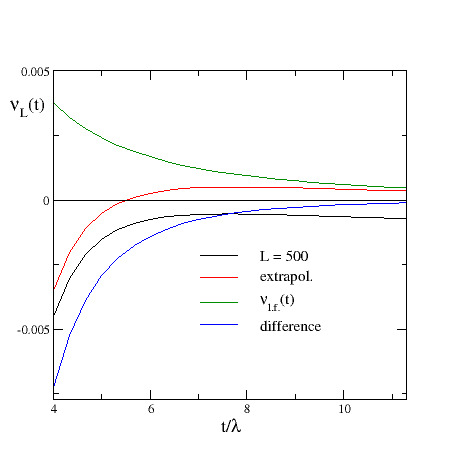,width=7cm,height=4cm}}
\caption{\label{fig:num3}(Color online) Same as right panel of
Fig.~\ref{fig:num1} for several variational parameters, {\it cf.}
Eq.~(\ref{eq:profile}). Top row: $\beta=1$ with $\alpha=0.85$ (left panel) 
and $\alpha=1.0$ (right panel). Bottom row: $\alpha=0.85$ with $\beta=0.5$ 
(left panel) and $\beta=1.5$ (right panel). The horizontal axis in each case 
is expressed in units of $\lambda = m/\sqrt{2}$.}
\end{figure}

We show the asymptotic behavior for different profiles in Fig.~\ref{fig:num3}.
The behavior determined by the ultra-violet analysis is well hidden
when the scalar potential is very attractive, either by being wide
(small $\alpha$) or deep (large $\beta$).  The weaker the scalar
potential, the more closely the extrapolation of $\nu_{L}(t)$ follows
the limiting function. For none of the considered cases did we
observe that the large but fixed $L$ result matched the limiting function. 

The strong enhancement at small $t$ is well fitted by the integrable
function $A_1+A_2{\rm ln}(\frac{t-\mu}{\lambda})$ with constants
$A_{1,2}$ and $\mu\lesssim m$. We associate this behavior with the second
order Born terms that are subtracted in $\nu_L(t)$ for the scalar
potential. It is thus not surprising (and actually is required) that 
we observe a similar behavior for $\overline{\nu}^{(2)}(t)$, as
defined in Eq.~(\ref{eq:lf5}). When computing the VPE, we solve 
the relevant differential equations starting at $t=1.45 \lambda$ 
and fit such a logarithmic function directly to the integrand of 
Eq.~(\ref{eq:almost_finished}) to obtain the contribution from the 
interval $t/\lambda\in[\sqrt{2},1.45]$. That interval typically contributes 
about 5\% to the full integral. Finally, we are in the position to give 
numerical results for the VPE in table~\ref{tab:VPE}.
\begin{table}
\centerline{\begin{tabular}{c|c||c|c|c||c}
$\alpha$ & $\beta$ & 
$\displaystyle E^{\rm scat.}_{\rm VPE}/\lambda^2$ 
& $\displaystyle \overline{E}^{(S)}_{\rm VPE}/\lambda^2$ 
& $\displaystyle  E_{\rm VPE}/\lambda^2$ 
& $\displaystyle \overline{E}^{(A)}_{\rm VPE}/\lambda^2$ \cr
\hline
0.50 & 1.0 & -0.0862 &  0.0061  & -0.0801 & -0.0017 \cr
0.85 & 1.0 & -0.0184 &  0.0053  & -0.0131 & -0.0051 \cr
1.00 & 1.0 & -0.0089 &  0.0050  & -0.0039 & -0.0072 \cr
0.85 & 1.5 & -0.0917 &  0.0127  & -0.0790 & -0.0051 \cr
0.85 & 0.5 & -0.0002 &  0.0010  &  0.0008 & -0.0051 \cr 
\end{tabular}}
\caption{\label{tab:VPE}The vacuum polarization energy (VPE) 
per unit length (in units of $\lambda^2=m^2/2$) and its components 
for several variational parameters, {\it cf.} 
Eq.~(\ref{eq:profile}) in the $\overline{\rm MS}$ renormalization
scheme. Also presented is the second order gauge field contribution
of Eq.~(\ref{eq:EAmsbar2}).}
\end{table}

Though the main purpose of this calculation has been to show the feasibility 
of this systematic approach, we also see that the VPE varies strongly with 
the strength of the scalar potential, which dominates the VPE. This result
is corroborated by the small contribution of the renormalized second
order Feynman diagrams of Fig.~\ref{fig:Qdiv}a) and b) listed in the
last column of table~\ref{tab:VPE}.

\section{Summary and outlook}

We have resolved a number of technical subtleties that are encountered
when applying the efficient spectral methods to the computation of the
one-loop quantum correction to the energy per unit length of a
Abrikosov-Nielsen-Olesen vortex. Because of the singular (topological) structure
of this configuration, identifying the ultra-violet divergences that must be
renormalized is nontrivial. In particular, the standard approach of equating 
elements of the Born series for the scattering data with (dimensionally) 
regularized Feynman diagrams fails because neither is well-defined. 
This problem is confined to the coupling of the vector potential to the 
scalar (Higgs) quantum fluctuations, which we have therefore separated
out in this study. As a first step, we considered a regular vector potential, 
and used the standard formulation of the spectral methods to verify that 
we then obtain the expected divergences. In doing so, we have identified 
a constant that must be subtracted from the angular momentum sum of the 
logarithm of the Jost function to maintain its analytic properties. This 
constant corresponds to the quadratic divergence of the loop diagrams with 
two insertions of the vector potential, which only cancels in a gauge
invariant regularization. Hence this subtraction merely enforces gauge 
invariance. Since the quantum correction to the energy is an integral that 
contains the derivative of that sum, any constant may be subtracted without 
changing the result; we then integrate by parts for calculational convenience. 
We have then verified that no sub-leading logarithmic divergence is left over 
from these diagrams after subtraction of this constant other than the one
eliminated by the standard wave-function renormalization  of the gauge
field. For non-singular backgrounds, the behavior of the Jost function 
corresponding to this divergence sets in at relatively small momenta, 
for which the angular momentum sum converges reasonably fast. This 
situation changes dramatically for the singular background of the vortex. 
Not only is the onset of the correct asymptotic behavior shifted to larger 
momenta, depending on the strength of the scalar potential, but without 
proper extrapolation it may emerge from the sum over partial waves only when 
the sum is extended to extremely large values for which, unfortunately,
the numerical calculation becomes unstable. For the
generic example that we have studied exhaustively, no finite
truncation of the angular momentum sum produced the correct behavior
and an extrapolation was unavoidable. Furthermore, large angular momenta 
were required to obtain an accurate extrapolation. It may
well be that the numerical effect is small when not matching the
correct large momentum behavior and we could instead omit that
part of the momentum integral when computing the vacuum polarization
energy. Even if so, that can only be justified {\it a posteriori} and
matching the asymptotics is essential for consistently renormalizing
the quantum energy.

With these technical obstacles solved, the next step is to consider
the full Higgs-gauge boson model that also treats the
gauge fields quantum mechanically. For the transverse gauge modes
there will be a scalar potential similar to $V_H$ and a coupling to
the scalar quantum fluctuations by a vertex that connects to the
vector potential profile. The temporal and longitudinal modes will
only couple to the scalar potential. In three space dimensions,
a particular gauge can be chosen such that the contributions to the
quantum energy from the temporal and longitudinal modes cancel against
those from the ghost fields needed for gauge fixing
\cite{Lee:1994pm,Baacke:2008sq}, while in two dimensions this cancellation 
does not occur and these modes must be explicitly included.  Of course,
the full model calculation will also require physical on-shell
renormalization conditions. Their implementation requires the momentum
space analysis of Feynman diagrams beyond the ultra-violet
divergences. We have seen how to treat obstacles that in this context
arise from the Fourier transform of the singular vortex background by
expressing such Feynman diagrams as the Fourier transform of the
elements of the (regular) field strength tensor. This will allow us
to compute the VPE of a superconducting vortex \cite{Abrikosov:1956sx}.

Ultimately we will be able to compare the quantum corrections to classical 
configurations in different topological sectors by replacing 
$\varphi\,\to\,n\varphi$ and $f_G\,\to\,nf_G$ with integer $n$ in 
Eq.~(\ref{eq:vortex}). This calculation will provide further insight into 
the Skyrme model picture for nuclei, which estimates the nuclei binding 
energies as differences of classical energies in sectors whose topological 
charge equals the number of nucleons and energies due to canonical
quantization of rotational modes~\cite{Manko:2007pr}. With respect to
the $\hbar$ counting, these are the leading and
next-to-next-to-leading order contribution, while the next-to-leading
order, the vacuum polarization energy, is omitted. It does not
contribute to the mass differences within a given topological sector,
but may alter the picture when different sectors are compared.

It will also be feasible to apply the proposed method to
supersymmetric extension  of the vortex model because the techniques
to include fermions have already been
established~\cite{Graham:2004jb}. There are (mainly analytical)
results~\cite{Schmidt:1992cu,Lee:1994pm} based on mode number
counting. However, such arguments can require more thorough 
investigation when renormalization is needed~\cite{Graham:1998qq}.

\acknowledgments
N.\@ G.\@ is supported in part by the National Science Foundation (NSF)
through grant PHY-1820700.
H.\@ W.\@ is supported in part by the National Research Foundation of
South Africa (NRF) by grant~109497.

\end{document}